\documentclass[11pt,twoside]{article}

\usepackage{asp2006}
\usepackage{epsf}
\usepackage{psfig}
\usepackage{lscape}
\usepackage{natbib}

\begin{document}

\title{Multi-Aperture Imaging of Extrasolar Planetary Systems}

\author{O. Absil}   

\affil{D{\'e}partement\ d'Astrophysique, G\'eophysique \& Oc\'eanographie, Universit\'e de Li\`ege, 17 All\'ee du Six Ao\^ut, B-4000 Sart Tilman, Belgium}

\begin{abstract}
In this paper, we review the various ways in which an infrared stellar interferometer can be used to perform direct detection of extrasolar planetary systems. We first review the techniques based on classical stellar interferometry, where (complex) visibilities are measured, and then describe how higher dynamic ranges can be achieved with nulling interferometry. The application of nulling interferometry to the study of exozodiacal discs and extrasolar planets is then discussed and illustrated with a few examples.
\end{abstract}


\section{Introduction}

When considering the direct imaging of extrasolar planetary systems, one is faced with two main challenges: the small angular separation and the high contrast between exoplanets and their host stars. If the goal is to characterise the mid-infrared emission of exoplanets in the habitable zone of nearby stars, the typical angular resolution of 50\,mas required to resolve a Sun-Earth system at 20\,pc leads to an impractical aperture size of 40\,m. The only option is then stellar interferometry, which synthesises the resolving power of a larger aperture by using multiple telescopes separated by an appropriate distance called the interferometric baseline $B$. The associated angular resolution equals $\lambda/2B$ (to be compare with $\lambda/D$ for a single aperture of diameter $D$). The need for interferometry to image planetary systems in the infrared was recognised already in the late 70s \citep{Bracewell78}. Its application to the search and characterisation of habitable worlds was proposed 15~years later by \citet{Leger93}.

Reaching the appropriate angular resolution to separate the signals of the planet from that of its host star is only half of the solution, and the high dynamic range still needs to be addressed. In this review, we discuss the various techniques that can be implemented with infrared interferometers to reach the required dynamic range to image extrasolar planetary systems of various kinds (from dusty discs and hot giant planets down to Earth-like planets).


\section{Classical Stellar Interferometry}

The main quantity that is measured by a (two-telescope) stellar interferometer is the complex visibility of the interference fringes recorded at the detector. The complex visibility of a high-contrast binary system, given by the sum of the complex visibilities of the two objects, can be approximated as follows:
\begin{equation}
{\cal V} \simeq (1-r){\cal V}_{\star} + r\exp(i2\pi(u\Delta\alpha+v\Delta\beta)) \;
\end{equation}
with ${\cal V}_{\star}$ the complex visibility of the central star, $u$ and $v$ the Fourier plane coordinates associated to the baseline vector $\mathbf{B}$ ($u=B_x/\lambda$, $v=B_y/\lambda$), $r$ the flux ratio between the planet and the star, and $\alpha$ and $\beta$ the cartesian angular coordinates of the planet relative to the star in the sky plane. In this expression, we have assumed that $r\ll 1$ and that the amplitude of the complex visibility of the sole planet is equal to $1$ (completely unresolved object). This expression shows that the presence of a faint companion around the target star affects both the amplitude $V=|{\cal V}|$ and the phase $\phi=\arg({\cal V})$ of the interference fringes, as follows:
\begin{eqnarray}
V^2 & \simeq & V_{\star}^2 - 2r(V_{\star}-\cos(2\pi(u\Delta\alpha+v\Delta\beta)) \\
\phi & \simeq & \frac{r\sin(2\pi(u\Delta\alpha+v\Delta\beta))}{V_{\star}}
\end{eqnarray}
where we have assumed that the star is point-symmetric (so that ${\cal V}_{\star}$ is real). In both cases, the magnitude of the planet-related effect is directly proportional to the flux ratio between the planet and the star.

With the existing, state-of-the-art ground-based interferometers, the typical accuracy that can be reached on the squared visibility is of the order of $10^{-2}$ on bright unresolved stars \citep[see e.g.,][]{Kervella04}. This level of performance does not allow extrasolar planets to be detected, even in the most favourable cases. Therefore, planet detection techniques mostly rely on phase measurements. The main problem with interferometric phases is that they are corrupted by atmospheric turbulence: the random piston between two apertures induces a time-varying phase that adds to the astrophysical phase. Furthermore, making an absolute measurement of the fringe phase would require a perfect knowledge of the zero optical path difference position in the instrument, which is impractical. Several techniques can however be used to retrieve (part of) the phase information. The main three of them are described below.

    \subsection{Wavelength-differential Phase}

When dispersing the interferometric fringes into several spectral channels, the fringe phase can be measured in one channel with respect to any other channel. Since the various wavelengths have travelled through the same atmosphere and through the same optical train, their phases should be the same unless the astrophysical object has a wavelength-dependent phase. In the case of a star-planet system, the phase is proportional to the flux ratio $r$, which changes with wavelength since the two bodies have different temperatures and various spectral signatures. By measuring the phase variations across the spectrum, one can thus theoretically reconstruct the variations of the star-planet contrast relative to a reference wavelength.

This method, which was thoroughly described by \citet{Vannier06}, is however not immune to chromatic effects in the Earth atmosphere and in the instrument. In particular, the compensation of the delay between the beams collected by two telescopes is generally done in the ambient (moist) air, whose refraction index changes with wavelength, thereby creating a wavelength-dependent phase. The magnitude of this effect can be predicted if the refraction index of air is modelled with a good enough accuracy. However, the fluctuations of the column density of water vapour (``water vapour seeing'') above the two telescopes creates a random chromatic component in the phase difference between the two beams, which is much more difficult to correct.

The most advanced use of this technique for extrasolar planet detection has been performed with the AMBER instrument at the VLTI \citep[see e.g.,][]{Millour08}. In practice, chromatic effects generally prevent the different phase to be measured with an accuracy better than a few 0.01\,radians, even when using advanced calibration techniques such as beam commutation. Flux ratios larger than 100:1 are therefore currently out of reach.

    \subsection{Phase Referencing}

Phase referencing consists in observing simultaneously two stars located within the same isoplanatic patch so that their phases can be measured relative to each other. This requires a high precision internal metrology system in order to monitor the non-common path differences between the two detected fringe packets. The final result is an accurate measurement of the relative positions of the two objects on the sky plane at the observing wavelength. If one of the two stars is accompanied by an exoplanet, the photocentre of the system will be slightly shifted towards the planet. Such a shift could potentially be detected, if the position of the sole star was known with a high enough accuracy. This is generally not the case, and phase referencing is therefore not appropriate to directly image an extrasolar planet. On the other hand, by measuring the time evolution of the position of a star with respect to a fixed object, one can evaluate the astrometric shift due the gravitational influence of planets around this star. This method has already been used at the Palomar Testbed Interferometer to search for planets in close binary system \citep{Lane04} and will soon be used at the VLTI with the PRIMA instrument \citep{Launhardt08}.

    \subsection{Closure Phase}

When using three telescopes (or more) at a time, one can exploit the closure phase $\Phi$, which is defined as the argument of the triple product of the complex visibilities measured on a closed triangle of baselines:
\begin{eqnarray}
\Phi & = & \arg(V_{12}e^{i\phi_{12}}V_{23}e^{i\phi_{23}}V_{31}e^{i\phi_{31}}) \\
     & = & \phi_{12} + \phi_{23} + \phi_{31}
\end{eqnarray}
By construction, this quantity is insensitive to telescope-specific phases, because an additional phase $\varphi_i$ on a given telescope $i$ appears positively in \linebreak\mbox{$\phi_{ij}=(\phi_i+\varphi_i)-\phi_j$} and negatively in $\phi_{ki}=\phi_k-(\phi_i+\varphi_i)$. The intrinsic closure phase of an astrophysical object can thus be measured by adding the phases of the fringes detected on the three baselines of a closed triangle, without being affected by atmospheric or instrument effects (including the chromatic effects that spoil differential phase measurements). Furthermore, one can show that the closure phase differs from zero only for targets that are not point-symmetric, which is precisely the case of high contrast binary systems. In this case, the closure phase can be written as follows assuming the host star to be unresolved:
\begin{equation}
\Phi \simeq r(\sin(2\pi\mathbf{u_{12}}\cdot\mathbf{\Delta}) + \sin(2\pi\mathbf{u_{23}}\cdot\mathbf{\Delta}) + \sin(2\pi\mathbf{u_{31}}\cdot\mathbf{\Delta}))\;
\end{equation}
with $\mathbf{u_{ij}}=(u_{ij},v_{ij})$ and $\mathbf{\Delta}=(\Delta\alpha,\Delta\beta)$. The closure phase is thus proportional to the flux ratio $r$, and its sensitivity to faint companions can potentially be significantly increased if the host star is resolved by at least one baseline of the triangle, as described by \citet{Chelli09}. Exoplanet host stars are however generally too small to be resolved with the current generation of interferometers.

Closure phase measurements are nowadays performed routinely by several interferometers around the world. Attempts to detect extrasolar planets with this method have been carried out with the MIRC instrument at the CHARA array \citep{Zhao08}, and with the AMBER instrument at the VLTI. In the best cases, the closure phase accuracy reaches $\sim 0.1^{\circ}$, allowing low-mass companions with contrast up to 600:1 to be detected at $1\sigma$. It is expected that improvements in the accuracy of closure phase measurements could soon provide the first direct detection of a hot extrasolar giant planet with interferometry. The combination of spectral dispersion with closure phase measurements (as done in AMBER) could potentially allow the detection of molecular bands in the exoplanet's atmosphere, such as CH$_4$ or CO features in the infrared $K$~band.


\section{Nulling Interferometry}

\begin{figure}[!ht]
\plottwo{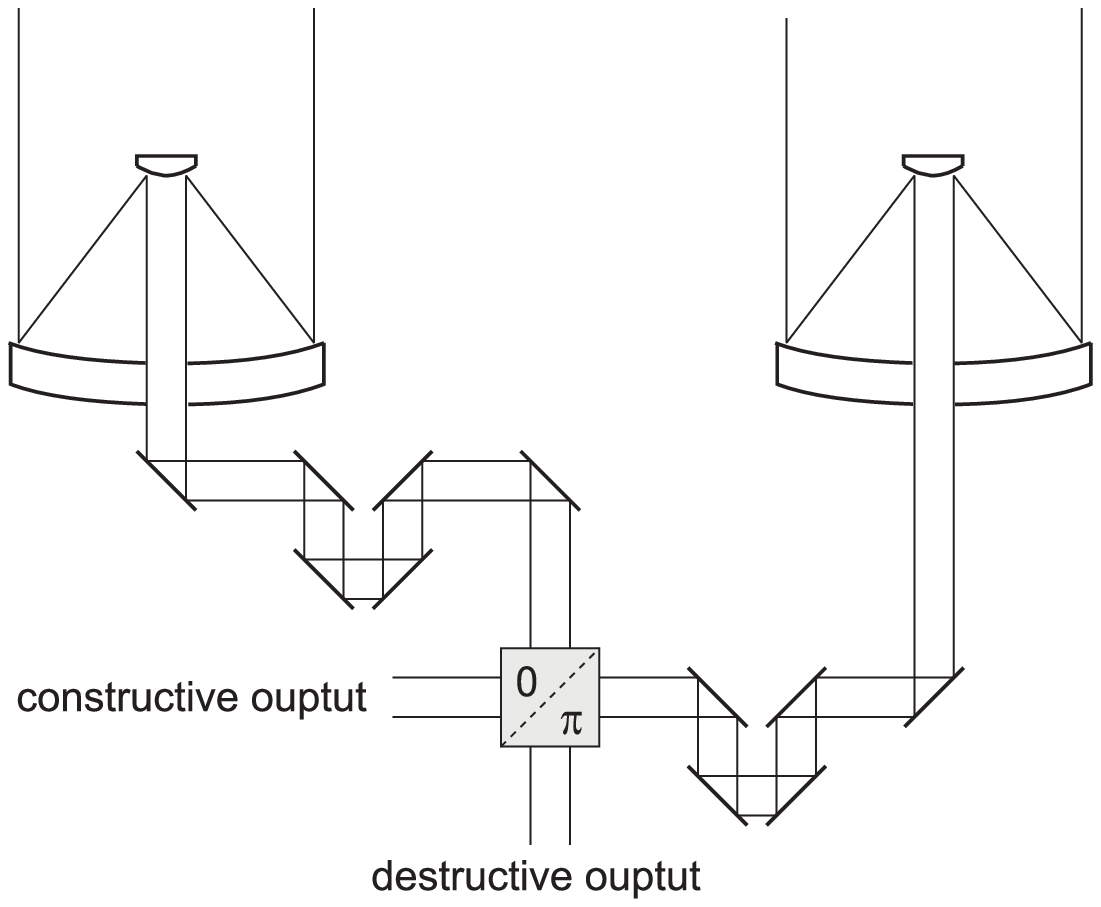}{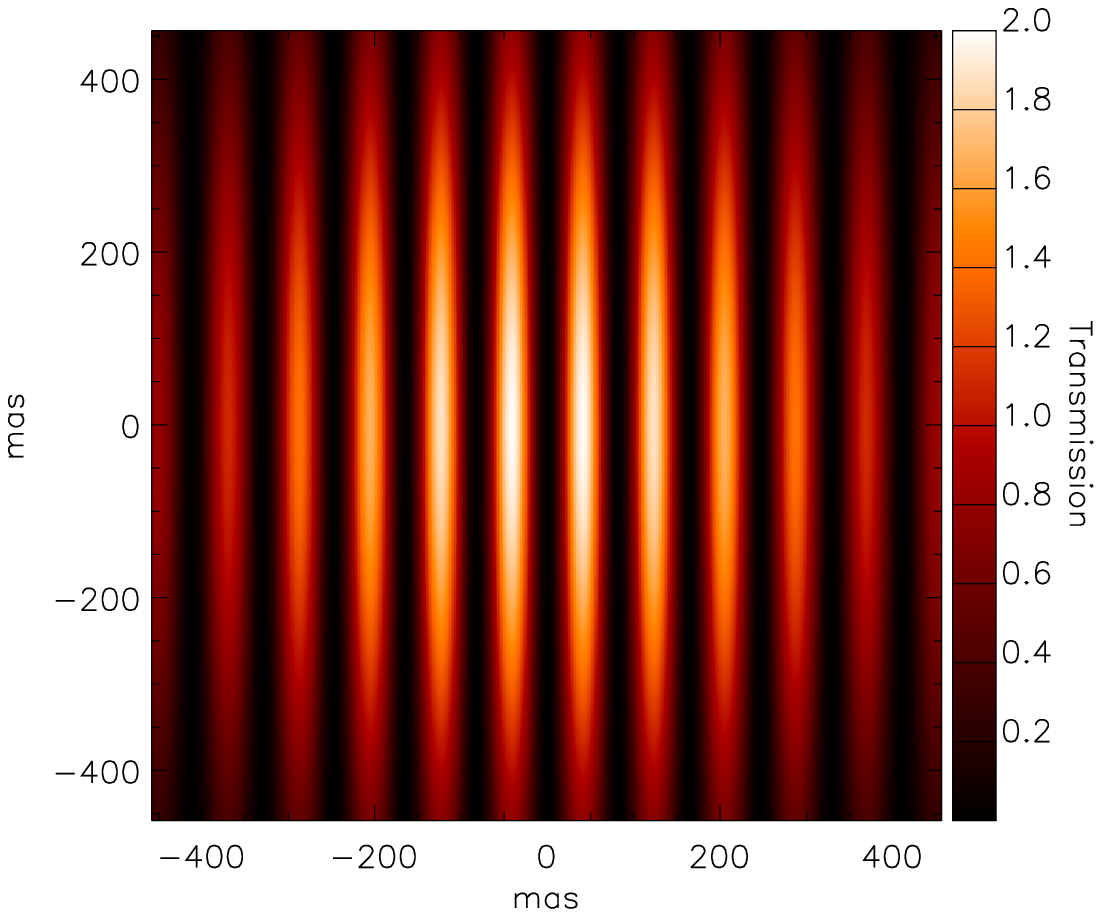}
\caption{\textit{Left}: Principle of a two-telescope nulling interferometer. The beam-combining system produces a destructive interference by applying a $\pi$ phase shift to one of the two input beams and by superposing them in a coaxial way. \textit{Right}: Associated transmission map, showing the parts of the field that are transmitted (bright stripes) and those that are blocked (dark stripes, including the central dark fringe) by the interference process.} \label{fig:brac}
\end{figure}

The main limitation to the capabilities of stellar interferometry in the detection of extrasolar planets resides in the high dynamic range that must be achieved. The previous section has shown that contrasts of 1000:1 and higher are difficult to reach with the current generation of ground-based arrays. One avenue to improve the dynamic range of interferometers is to use the undulatory nature of light to perform a destructive interference on the blinding stellar light. When overlapping the light beams collected by two telescopes on a balanced beam splitter, one can tune their respective phases by means of phase-shifting devices so that all the on-axis stellar light is sent to only one of the two complementary outputs of the beam splitter (see Fig.~\ref{fig:brac}). Such a configuration happens when a phase shift of $\pi$~radians is maintained between the two input beams. This is in contrast with classical interferometry, where the optical path difference between the beams takes a range of values (either in a pupil plane or an image plane) in order to record a complete interferogram, or at least a part of it.

A two-telescope nulling interferometer is characterised by its transmission map $T_{\lambda}(\theta,\phi)$, displayed in Fig.~\ref{fig:brac}, which results from the fringe pattern produced by the interference between the two beams:
\begin{equation}
T_{\lambda}(\theta,\alpha) = 2 P_{\lambda}(\theta,\alpha) \sin^2\left(\pi \frac{B\theta}{\lambda} \cos\alpha \right) \:
\end{equation}
where $\theta$ and $\alpha$ are respectively the radial and polar angular coordinates with respect to the optical axis, $B$ the interferometer baseline (whose direction defines $\alpha=0$), $D$ the telescope diameter, $\lambda$ the wavelength, and where $P_{\lambda}(\theta,\phi)$ represents a field-of-view taper function resulting from the size of the collecting apertures and from the particular design of the instrument. For small values of $\theta$ ($\ll \lambda/B$), one can see that the interferometer transmission is proportional to $\theta^2$, so that the central part of the dark fringe is parabolic.

The final detection can be done either in a pupil or in an image plane. In the former case, a single-pixel detector is sufficient to record the total flux in the output pupil, emanating from all the sources in the diffraction-limited field-of-view. In the latter, an image similar to that of a single telescope is formed, except that the relative contribution of each source is affected by the interferometer's intensity response at its location. In any case, no fringes are formed, and the final output generally consists in a single value: the total intensity in the diffraction-limited field-of-view. The final output flux then writes:
\begin{equation} F(\lambda) = \int_{\alpha} \int_{\theta} 2 P_{\lambda}(\theta,\alpha) \left( \sin^2(\pi \frac{B\theta}{\lambda} \cos\alpha ) I_c(\theta, \alpha, \lambda) + I_i(\theta,\alpha,\lambda) \right) \, \theta \, {\rm d}\theta \, {\rm d}\alpha \; \label{eq:output}
\end{equation}
with $I_c(\theta, \alpha, \lambda)$ and $I_i(\theta, \alpha, \lambda)$ the intensity distributions of the coherent and incoherent sources located within the fied-of-view. The nulling ratio $N$ is then defined as the fraction of stellar light that makes it to the destructive output due to its finite angular diameter $\theta_{\star}$. Assuming the stellar angular diameter to be small compared to the angular resolution $\lambda/B$ of the interferometer, it can be shown that the nulling ratio reduces to:
\begin{equation}
N \simeq \frac{\pi^2}{16} \left(\frac{B\theta_{\star}}{\lambda} \right) ^2 \;  \label{eq:null}
\end{equation}

In principle, nulling interferometry can be generalised to any number of telescopes, provided that the phases introduced in each beam result in a destructive interference on the optical axis of the interferometer (i.e., for a zero optical path difference). In particular, using more that two telescopes at the same time allows deeper nulls to be achieved, with the central transmission proportional to higher powers of $\theta$ than the $\theta^2$ proposed by a two-telescope interferometer.

    \subsection{Two-telescope Nullers: Exozodi Finders}

Because all the sources located within the coherent field-of-view of a single aperture contribute (with various transmission factors) to the destructive output of the interferometer, a simple two-telescope nulling interferometer can only detect the sources that are producing the largest contribution. Stellar leakage, related to the finite size of the stellar photosphere, is generally one of the main contributors. However, its expression is analytical (Eq.~\ref{eq:null}), so that its contribution can be subtracted \textit{a posteriori} if the stellar diameter $\theta_{\star}$ is known with a good enough accuracy. Another major contribution to the flux detected at the destructive output is the incoherent background, produced by the thermal emission of the Earth atmosphere and of the instrument itself. Since this emission is uniform across the field-of-view, its contribution can be suppressed e.g.\ by rotating the baseline of the interferometer and subtracting two measurements.

Once the star and background contributions have been subtracted, the remaining contribution comes from the immediate environment of the star, i.e., the planetary system in our case. The largest source of infrared emission in extrasolar planetary systems generally comes from the cloud of interplanetary dust, which represents the primordial material from which planets are formed in the case of protoplanetary discs, or results from the collisional activity of larger bodies in the case of exozodiacal discs (the extrasolar equivalent of the zodiacal disc of dust grains surrounding our Sun). Owing to their brightness, such discs are generally the main targets of two-telescope nullers. Hot giant planets could also be detected in a few cases.

\begin{figure}
\plotone{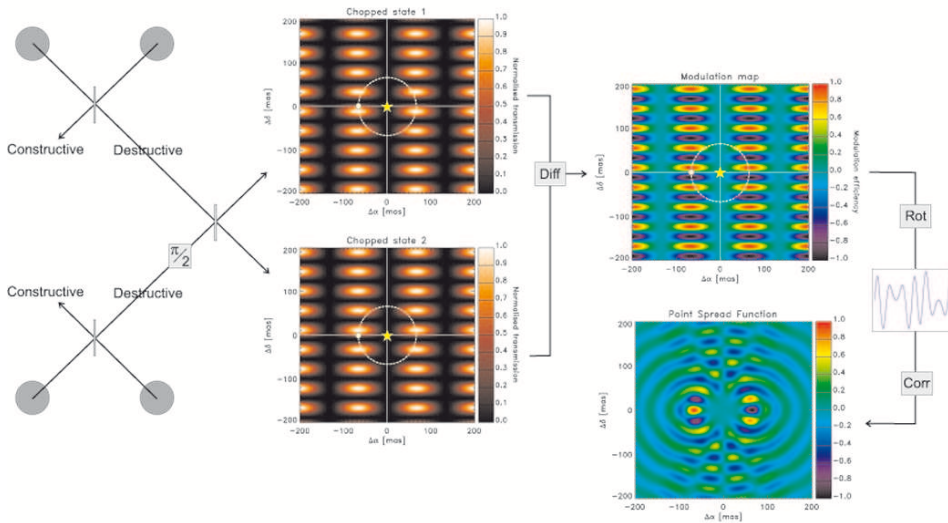}
\caption{Principle of phase chopping, illustrated for the X-array configuration. Combining the beams with different phases produces two conjugated transmission maps (or chop states), which are used to produce the chopped response. Array rotation then locates the planet by cross-correlation of the modulated chopped signal with a template function.} \label{fig:chopping}
\end{figure}

Until now, only two ground-based nulling instruments have been producing scientific results, in particular in the field of extrasolar planetary systems. The BLINC nulling camera has been used since 1998 with two sub-apertures of the 6.5-m MMT at Mount Hopkins, Arizona. The observations of bright main sequence stars in the $N$ band have resulted in $3\sigma$ upper limits on the dust density of exozodiacal discs ranging between 220 and 10,000~zodi \citep[1~zodi being the luminosity of our own zodiacal cloud as seen from outside the solar system, see][]{Liu09}. The second nulling instrument is the Keck Interferometer Nuller \citep[KIN,][]{Colavita09}, which is combining the lights of the two 10-m Keck telescopes in the $N$ band on top of Mauna Kea, Hawaii. In order to subtract the background from the scientific measurements, this instrument divides the pupils of the Keck telescopes into two halves, produces two (equivalent) nulled outputs on the 85-m Keck baseline using two pairs of half pupils, and then combines the two nulled outputs with a time-varying phase shift in order to modulate the coherent signal contributing to the nulled outputs. Synchronous demodulation can the be used to remove the incoherent background. This process is similar to the phase chopping technique discussed in the next section and illustrated in Fig.~\ref{fig:chopping}. KIN reaches a typical sensitivity of 300~zodi for nearby main sequence stars, and has been performing a survey for exozodiacal discs on a sample of about 40~stars. The first results are currently being published \citep[see first paper by][]{Stark09}.

\begin{figure}
\plotone{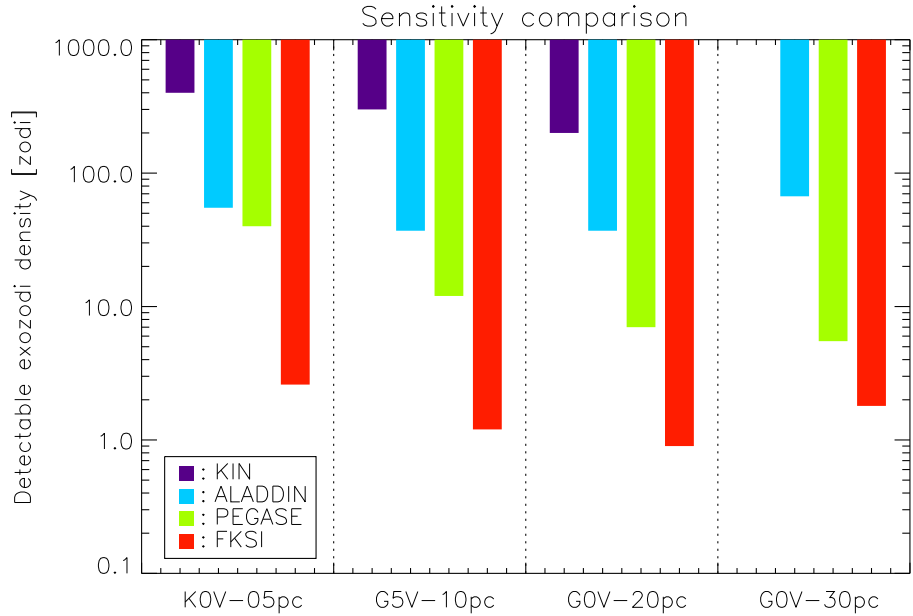}
\caption{Measured or simulated performance of various ground- and space-based nulling interferometers, in terms of the smallest exozodiacal dust density that can be detected at $3\sigma$ around various solar-type stars.} \label{fig:perfo}
\end{figure}

Several projects of ground-based or space-based two-telescope nulling interferometers are currently being considered. The ALADDIN project \citep{Absil07} aims at performing deep nulling in the $L$ band on the high Antarctic plateau, where the atmospheric conditions (turbulence, background, transparency) are well suited to such observations. Pegase and FKSI \citep{Defrere08} are two space mission projects with similar goals and designs, except that Pegase is based on a free flying flotilla of three spacecrafts, while FKSI relies on a structurally connected interferometer with a 12-m baseline. The expected performance of these projects are compared to the KIN sensitivity in Fig.~\ref{fig:perfo}.

    \subsection{Multi-telescope Nullers: Planet Imagers}

Because planets (and in particular Earth-like planets) are generally not the brightest component in extrasolar planetary systems, a simple two-telescope nulling interferometer is mostly inappropriate for their detection. An additional subtraction technique must be used to get rid of the exozodiacal disc. A few techniques have been considered so far, mostly relying on the fact that the intensity distribution of the exozodi is point-symmetric while the planet is an off-axis point-like source. The most promising of these techniques is phase chopping. Its principle is to synthesise two different transmission maps with the same telescope array, by applying different phase shifts in the beam combination process. In order to remove all point-symmetric sources in the field-of-view (star, background, exozodiacal disk) while keeping the planetary signal, the two maps must be linked to each other by point symmetry, but have no point symmetry themselves. This phase chopping technique can be implemented in various ways, and is now an essential part of the future space-based planet finding missions such as the Darwin and TPF-I projects that have been considered by ESA and NASA respectively during the last decade. Based on a free-flying array of four 2-m class telescopes, these missions are capable of detecting and spectroscopically characterising the thermal emission from Earth-like planets located in the habitable zone of nearby solar-type stars. The interested reader is referred to the recent review by \citet{Cockell09} for a detailed description of these missions, and to \citet{Defrere10} for a description of the mission performance and of the influence of exozodiacal light on this performance.

\acknowledgements 

The author acknowledges the financial support from an F.R.S.-FNRS postdoctoral fellowship, and from the Communaut\'e Fran{\c c}aise de Belgique (ARC -- Acad\'emie universitaire Wallonie-Europe).



\begin{thebibliography}{}
\bibitem[Absil et al.(2007)]{Absil07}Absil, O., Coud{\'e} du Foresto, V., Barillot, M., \& Swain, M. 2007, \aap, 475, 1185
\bibitem[Bracewell(1978)]{Bracewell78}Bracewell, R. 1978, \nat, 274, 780
\bibitem[Chelli et al.(2009)]{Chelli09}Chelli, A., Duvert, G., Malbet, F., \& Kern, P. 2009, \aap, 498, 321
\bibitem[Cockell et al.(2009)]{Cockell09}Cockell, C., Herbst, T., L{\'e}ger, A.,  et al. 2009, Exp. Astron., 23, 435
\bibitem[Colavita et al.(2009)]{Colavita09}Colavita, M., Serabyn, E., Millan-Gabet, R.,  et al.  2009, \pasp, 121, 1120
\bibitem[{Defr{\`e}re} et~al.(2008)]{Defrere08}{Defr{\`e}re}, D., {Absil}, O., Coud{\'e} du Foresto, V., Danchi, W., \& den Hartog, R. 2008, \aap, 490, 435
\bibitem[{Defr{\`e}re} et~al.(2010)]{Defrere10}{Defr{\`e}re}, D., {Absil}, O., {den Hartog}, R., {Hanot}, C., \& {Stark}, C. 2010, \aap, 509, A9
\bibitem[Kervella et al.(2004)]{Kervella04}Kervella, P., S{\'e}gransan, D., \& Coud{\'e} du Foresto, V. 2004, \aap, 425, 1161
\bibitem[Lane et al.(2004)]{Lane04}Lane, B. \& Muterspaugh, M. 2004, \apj, 601, 1129
\bibitem[Launhardt et al.(2008)]{Launhardt08}Launhardt, R., Queloz, D., Henning, T., et al.  2008, Proc. SPIE, 7013, 2I
\bibitem[L{\'e}ger et al.(1993)]{Leger93}L{\'e}ger, A., Mariotti, J.-M., Mennesson, B., et al. 1996, Icarus, 123, 249
\bibitem[Liu et al.(2009)]{Liu09}Liu, W., Hinz, P., Hoffman, W., Brusa, G., Miller, D., \& Kenworthy, A. 2009, \apj, 693, 1500
\bibitem[Millour et al.(2008)]{Millour08}Millour, F., Petrov, R., Vannier, M., \& Kraus, S. 2008, Proc. SPIE, 7013, 1G
\bibitem[Stark et al.(2009)]{Stark09}Stark, C., Kuchner, M., Traub, W., Monnier, J. D., et al. 2009, \apj, 703, 1188
\bibitem[Vannier et al.(2006)]{Vannier06}Vannier, M., Petrov, R., Lopez, B., \& Millour, F. 2006, \mnras, 367, 825
\bibitem[Zhao et al.(2008)]{Zhao08}{Zhao}, M., {Monnier}, J.~D., {ten Brummelaar}, T., {Pedretti}, E., \& {Thureau}, N.~D. 2008, Proc. SPIE, 7013, 1K
\end{thebibliography}

\end{document}